\documentclass[a4paper,12pt]{article}
\usepackage{pos}
\usepackage{braket}

\title{Determination of quark and lepton masses and mixings in the microscopic model}

\author{Bodo Lampe}

\affiliation{II. Institut für Theoretische Physik\\ Universität Hamburg, Germany}

\emailAdd{lampe.bodo@web.de}

\abstract{Recently, formulas for the mixing matrices of quarks and leptons have been put forward. My contribution here describes the relevant foundational and technical aspects which have led to those results.\\
The work has been carried out in the framework of the microscopic model\cite{bodomasses}. The most general ansatz for the interactions among tetrons leads to a Hamiltonian $H$ involving Dzyaloshinskii-Moriya (DM), Heisenberg and torsional isospin forces. Diagonalization of the Hamiltonian provides for 24 eigenvalues which are identified as the quark and lepton masses. While the masses of the third and second family arise from DM and Heisenberg type of isospin interactions, light family masses are related to torsional interactions among tetrons. Neutrino masses turn out to be special in that they are given in terms of tiny isospin non-conserving DM, Heisenberg and torsional couplings.\\ 
The approach not only leads to masses, but also allows to calculate the quark and lepton eigenstates, an issue, which is important for the determination of the CKM and PMNS mixing matrices. The almost exact isospin conservation of the system dictates the form of the lepton states and makes them independent of all the couplings in $H$. Much in contrast, there is a strong dependence of the quark states on the coupling strengths, and a promising hierarchy between the quark family mixings shows up.}

\FullConference{The European Physical Society Conference on High Energy Physics (EPS-HEP2023)\\
 21-25 August 2023\\
Hamburg, Germany\\}

\begin{document}
\maketitle


In the microscopic model\cite{bodomasses} quarks and leptons arise as eigenmode excitations of an internal tetrahedral fiber structure, which is made up from 4 constituents and extends into 3 extra dimensions. The constituents are called tetrons and transform under the fundamental spinor representation 8 of SO(6,1). 

More in detail, the ground state of the model looks like illustrated in Fig. 1. Each tetrahedron is made up from 4 tetrons, depicted as dots. The picture is a little misleading because physical space and the extra dimensions are assumed to be completely orthogonal.

With respect to the decomposition of $SO(6,1)\rightarrow SO(3,1)\times SO(3)$ into the (3+1)-dimensional base space and the 3-dimensional internal space, a tetron $\Psi$ possesses spin $\frac{1}{2}$ and isospin $\frac{1}{2}$. This means it can rotate both in physical space and in the extra dimensions, and corresponds to the fact that $\Psi$ decomposes into an isospin doublet $\Psi=(U,D)$ of two ordinary SO(3,1) Dirac fields U and D.
\begin{eqnarray}
8 \rightarrow (1,2,2)+(2,1,2)=((1,2)+(2,1),2)
\label{eq8}
\end{eqnarray}
For the $\Psi$ field left and right handed `isospin vectors' may be defined
\begin{eqnarray}
\vec Q_L=\frac{1}{4}\Psi^\dagger (1-\gamma_5)\vec \tau\Psi =\frac{1}{2}\Psi^\dagger_L \vec\tau \Psi_L
\qquad \quad
\vec Q_R= \frac{1}{4}\Psi^\dagger (1+\gamma_5)\vec \tau\Psi =\frac{1}{2}\Psi^\dagger_R \vec\tau \Psi_R 
\label{eq894}
\end{eqnarray}
as well as the corresponding densities
\begin{eqnarray}
n_L=\frac{1}{4}\Psi^\dagger (1-\gamma_5)\Psi =\frac{1}{2}\Psi^\dagger_L  \Psi_L
\qquad \quad
n_R= \frac{1}{4}\Psi^\dagger (1+\gamma_5)\Psi =\frac{1}{2}\Psi^\dagger_R \Psi_R
\label{pm11gg92}
\end{eqnarray}
$\vec \tau=(\tau_x,\tau_y,\tau_z)$ are the Pauli matrices in `internal' isospin space, whose coordinates will be denoted as $x$, $y$ and $z$.

\begin{figure}[h]
\begin{center}
\includegraphics[width=6.0in]{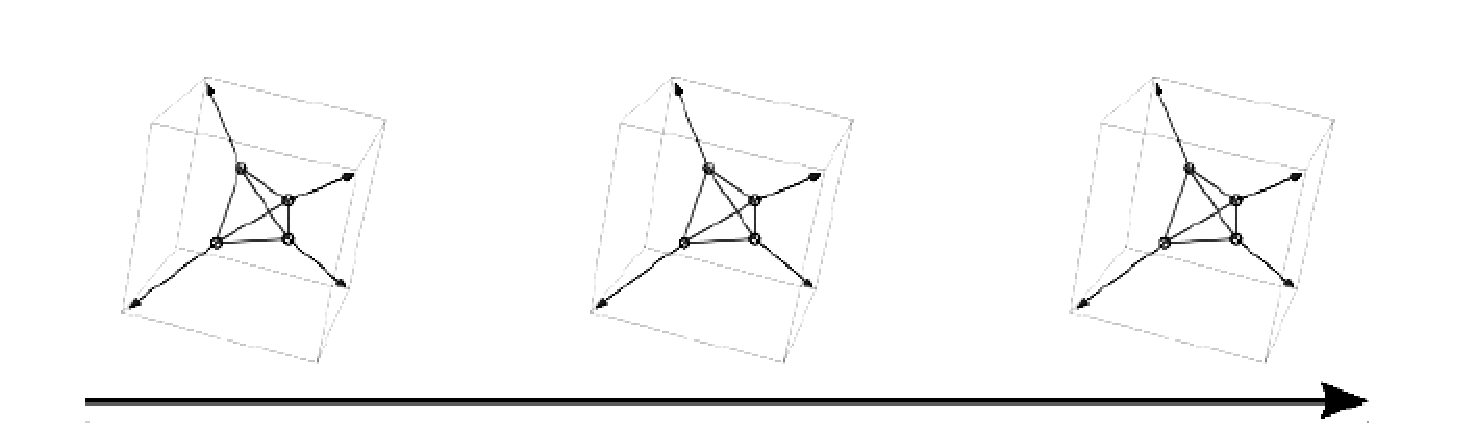}
\end{center}
\caption{The global ground state after the electroweak symmetry breaking has occurred, considered at Planck scale distances. The big black arrow represents 3-dimensional physical space. Before the symmetry breaking the isospin vectors are directed randomly, thus exhibiting a local SU(2) symmetry, but once the temperature drops below the Fermi scale $\Lambda_F$, they become ordered into a repetitive tetrahedral structure, thereby spontaneously breaking the initial SU(2). Note that the SM Higgs vev is related to the length of the aligned isospin vectors. Quarks and leptons glide on this background as quasiparticle excitations. The background has the properties of a Lorentz ether and is thereby not in conflict with Michelson-Morley type of experiments.}
\label{aba:fig1}
\end{figure}

The SM SSB being realized by an alignment of the tetron isospins, it is not surprising that the masses of quarks and leptons, and thus the SM Yukawa couplings are determined by the interactions among those isospins. The simplest interaction Hamiltonian between isospin vectors of 2 tetrons i and j is of the form $H=- J \,  \vec Q_{i}\vec Q_{j}$. 
So it has the form of a Heisenberg interaction - but for isospins, not for spins. The coupling J may be called an `isomagnetic exchange coupling'. 

\section{Relations between fermion masses and isospin couplings} 

In reality, the Hamiltonian H is more complicated, for several reasons:\\
$\bullet$ The appearance of antitetron degrees of freedom. This can be accounted for by using interactions both of $\vec Q_L$ and $\vec Q_R$
\begin{equation} 
H_H=-  \sum_{i\neq j=1}^4 \, [J_{LL}\,  \vec Q_{Li}\vec Q_{Lj}+ J_{LR}\,  \vec Q_{Li}\vec Q_{Rj}- J_{RR}\,  \vec Q_{Ri}\vec Q_{Rj}]
\label{mmHL3}
\end{equation}
for tetron fields located at tetrahedral sites  $i,j=1,2,3,4$, and it means that each vector in Fig. 1 should actually be interpreted as 2 vectors $\vec Q_L$ and $\vec Q_R$.\\ 
As seen below, the 3 couplings $J_{LL}$, $J_{LR}$ and $J_{RR}$ can be roughly associated to the quark and lepton masses of the second family.\\ 
$\bullet$ In addition to the Heisenberg Hamiltonian Dzyaloshinskii–Moriya interactions are to be considered. They will be shown to give the dominant mass contributions to the heavy family and are generically of the form
\begin{equation} 
H_{DM}=- K \,  \sum_{i\neq j=1}^4 \, \vec D_{ij} (\vec Q_{i} \times \vec Q_{j} ) 
\label{mm3444}
\end{equation}
The structure of the vectors $\vec D_{ij}$ is dictated by the tetrahedral symmetry to be $\vec D_{ij}=\vec Q_{i}\times \vec Q_{j}$\cite{moriya}.\\ 
$\bullet$ Heisenberg and DM terms do not contribute at all to the masses $m_e$, $m_u$ and $m_d$ of the first family. Therefore, small torsional interactions have to be introduced. They are characterized by the exerting torques $d\vec Q_{L,R}/dt$ being proportional to the isospins $\vec Q_{L,R}$ themselves.\\ 
$\bullet$ The masses of the neutrinos are yet another story. While the interactions discussed so far are isospin conserving and leave the neutrinos massless, neutrino masses can arise only from isospin violation\cite{bodotalk}. The treatment of the neutrino masses is not described here, but in a separate publication\cite{bodoprep}.

\vspace{0.2cm}
{\bf Dzyaloshinskii masses for the third family - Heisenberg masses for the second}

My presentation of the mass calculations begins with the Dzyaloshinskii-Moriya (DM) coupling, 
firstly because it is the dominant isospin interaction and secondly it gives masses only to 
the third family, i.e. to top, bottom and $\tau$, while leaving all other quarks and leptons massless. 
Among all the fermion masses the top quark mass is by far the largest and is of the order of the Fermi scale. As turns out, this is no accident, but has to do with the largeness of the relevant DM coupling. The complete DM Hamiltonian reads
\begin{eqnarray} 
H_D=- \sum_{i\neq j=1}^4 \, [K_{LL}  \,  (\vec Q_{Li} \times \vec Q_{Lj} ) ^2
- K_{LR} \,   (\vec Q_{Li} \times \vec Q_{Rj} ) ^2 
- K_{RR} \,    (\vec Q_{Ri} \times \vec Q_{Rj} ) ^2 ] 
\label{mmDM3}
\end{eqnarray}
with DM couplings $K_{LL}$, $K_{LR}$ and $K_{RR}$. 

It is convenient to already include at this point the Heisenberg terms (\ref{mmHL3}). They give masses both to the second and third family (but not to the first one) and their couplings J are typically smaller than 1 GeV, while the DM couplings K are larger. Altogether, Heisenberg and DM terms provide the most general isotropic and isospin conserving interactions within the internal space. Apart from that there will only be tiny torsional interactions responsible for the mass of the first family.

As envisaged, the quarks and leptons are vibrations $\vec \delta$ of the isospin vectors $\vec Q_{Li}$ and $ \vec Q_{Ri}$ of the tetrons $i$ at sites $i=1,2,3,4$, i.e. fluctuations of the ground state values of isospin vectors (\ref{eq894}) within one tetrahedron 
\begin{eqnarray}
\vec Q_{Li}=\langle \vec Q_{Li}\rangle+\, \vec \delta_{Li} \qquad \qquad \qquad \vec Q_{Ri}=\langle \vec Q_{Ri}\rangle+\, \vec \delta_{Ri}
\label{eqxxdrt1}
\end{eqnarray}
where $\langle \vec Q_{Li} \rangle$ and $\langle \vec Q_{Ri} \rangle$ 
are the ground state radial isospin vectors of a tetrahedron in Fig.1.

The masses of the excitations can be calculated by diagonalizing torque equations of the generic form
\begin{equation} 
\frac{d\vec Q}{dt} = i \, [H, \vec Q] 
\label{txxm32}
\end{equation}
and using the angular momentum commutation relations for the isospin vectors
\begin{equation} 
[Q_{Ri}^{a},Q_{Rj}^{b}]=i\delta_{ij}\epsilon^{abc} Q_{Ri}^{c} 
\qquad
[Q_{Li}^{a},Q_{Lj}^{b}]=i\delta_{ij}\epsilon^{abc} Q_{Li}^{c} 
\qquad
[Q_{Ri}^{a},Q_{Lj}^{b}]=0 
\label{mm3444b}
\end{equation}
where $i,j=1,2,3,4$ count the 4 tetrahedral edges and $a,b,c=1,2,3$ the 3 internal  directions(=extra dimensions). Note that while the masses correspond to the eigenvalues, CKM and PMNS mixings can be deduced from the eigenvectors. This point will be discussed in Appendix II.

The 24 first order differential equations for the $\vec \delta$ are rather lengthy. In linear approximation they read 
\begin{eqnarray} 
\frac{d\vec \delta_{Li}}{dt}&=&  2 K_{LL}  \{ \vec Q_{0}\times \vec \Delta_{LLi} +i [-\vec \Delta_{LLi} +(\vec \Delta_{LLi} .\vec Q_0) \, \vec Q_0]\} \nonumber \\
&+& 2 K_{LR}  \{ \vec Q_{0}\times \vec \Delta_{LRi} +i [-\vec \Delta_{LRi} +(\vec \Delta_{LRi} .\vec Q_0) \,\vec Q_0]\} \nonumber\\
&+& J_{LL}  ( \vec Q_{0}\times \vec \Delta_{LLi}) + J_{LR}  ( \vec Q_{0}\times \vec \Delta_{LLi}) 
\label{allnxxg} \\
\frac{d\vec \delta_{Ri}}{dt}&=&  2 K_{RR}  \{ \vec Q_{0}\times \vec \Delta_{RRi} +i [-\vec \Delta_{RRi} + (\vec \Delta_{RRi} .\vec Q_0) \,\vec Q_0]\} \nonumber \\
&+& 2 K_{LR}  \{ \vec Q_{0}\times \vec \Delta_{RLi} +i [-\vec \Delta_{RLi} + (\vec \Delta_{RLi} .\vec Q_0)\, \vec Q_0]\} \nonumber \\
&+& J_{RR}  ( \vec Q_{0}\times \vec \Delta_{RRi}) + J_{LR} (  \vec Q_{0}\times \vec \Delta_{RLi} )
\label{allnxxg1}
\end{eqnarray}
In these equations $\vec \delta_{Li}=\vec Q_{Li}- \langle \vec Q_{Li}\rangle $ and $\vec \delta_{Ri}=\vec Q_{Ri}- \langle \vec Q_{Ri}\rangle $,  $a=1,2,3,4$, denote the isospin vibrations and the $\Delta$'s are certain linear combinations of them which are important to maintain isospin conservation
\begin{eqnarray} 
\vec \Delta_{LLi}&=&- 3 \,\vec \delta_{Li} + \sum_{j\neq i} \vec \delta_{Lj}  \nonumber \\
\vec \Delta_{LRi}&=&- 3 \,\vec \delta_{Li} + \sum_{j\neq i} \vec \delta_{Rj}  \nonumber \\
\vec \Delta_{RLi}&=&- 3 \,\vec \delta_{Ri} + \sum_{j\neq i} \vec \delta_{Lj}  \nonumber \\
\vec \Delta_{RRi}&=&- 3 \,\vec \delta_{Ri} + \sum_{j\neq i} \vec \delta_{Ri}  
\label{anx5}
\end{eqnarray}
Eqs. (\ref{allnxxg}) and (\ref{allnxxg1}) correspond to a 24$\times$24 eigenvalue problem which - after the SSB - leads to 6 singlet and 6 triplet states, the latter ones each consisting of 3 degenerate eigenstates (corresponding to three quark colors).

After diagonalization one obtains the following results: the first family excitations are still massless at this point, but will get masses from the torsional interactions to be discussed below. The DM exchange coupling $K_{LL}$ is consistently of the order of the transition energy $\Lambda_F$ resp. the top quark mass, and the DM and Heisenberg couplings  can be accommodated to reproduce the third and second family masses. 

Namely, assuming the DM couplings K to dominate over the Heisenberg couplings J,  one can prove the following approximate relations  
\begin{eqnarray} 
m_t &=& 4 K_{LL} +O(J)\, \;\;\;\;    m_\tau=\frac{3}{2} K_{LR} +O(J)\,\;\; \;\;    m_b=4 K_{RR}  +O(J) \nonumber \\
m_c&=& J_{LL}\,\;\; \;\; \;\; \;\;    m_\mu=\frac{3}{2} J_{LR}\,\;\; \;\;\;\; \;\;     m_s=J_{RR}
\label{all36}
\end{eqnarray}
One concludes that in this approximation, the masses of quarks and leptons arise from different isospin interaction terms in (\ref{mmHL3}) and (\ref{mmDM3}), each mass associated essentially to one of the interactions.

\vspace{0.2cm}
{\bf Isospin conserving torsion and the masses of the first family}

It was seen above how the heaviness of the third family is related to large DM couplings. Afterwards masses of the quarks and leptons of the second family were obtained from Heisenberg exchange. It then remains to show how the small masses of the first family can be obtained from isospin conserving torsional interactions. Actually, torsional interactions give contributions to the masses of all families. However, since they are assumed to be small, the 2 heavy families remain 
dominated by DM and Heisenberg couplings, as given in (\ref{all36}).

The structure of torsional interactions is quite simple. Using the notation introduced in (\ref{anx5})  one has
\begin{eqnarray}
\frac{d\vec\delta_{Li}}{dt}=i C_{LL} \vec \Delta_{LLi}   + i C_{LR} \vec \Delta_{LRi} 
\qquad\qquad
\frac{d\vec\delta_{Ri}}{dt}=i C_{LR}  \vec\Delta_{RLi}   + i C_{RR}  \vec\Delta_{RRi} 
\label{v377}
\end{eqnarray}
with torsional couplings $C_{LL}$, $C_{LR}$ and $C_{RR}$. Since (\ref{v377}) gives the only mass contributions to the first family, the C-couplings can be chosen to accommodate the mass of the up quark, down quark and electron, respectively. Namely, one arrives at the mass formulas
\begin{eqnarray}
m_e&=&6 C_{LR}\\
m_u&=&-2C_{LL}+3C_{LR}+2C_{RR}-W_C\\
m_d&=&-2C_{LL}+3C_{LR}+2C_{RR}+W_C
\label{v37a7}
\end{eqnarray}
where 
\begin{eqnarray} 
W_C:= \sqrt{4(C_{LL}+C_{RR})^2+C_{LR}^2}
\label{allep36}
\end{eqnarray}
Then, using the phenomenological values
\begin{eqnarray}
m_e=0.51\; MeV \qquad \quad m_u=1.7 \;MeV \qquad \quad m_d=4.7\; MeV 
\label{v37b7}
\end{eqnarray}
one obtains
\begin{eqnarray}
C_{LR}=0.085\; MeV \qquad \quad  C_{LL}=1.13 \;MeV \qquad \quad C_{RR}=0.49 MeV
\label{v37c7}
\end{eqnarray}

\section{Supplementary material I: Mathematica program to calculate the quark and lepton masses and eigenstates} 

The following code allows to calculate quark and lepton masses and eigenstates, given the isospin couplings as defined in the main text. The resulting masses can be found at the bottom line of the program (in GeV). As can be seen, the program also generates reasonably small neutrino masses. The way to generate these masses goes beyond the scope of this talk and will be described in a separate publication\cite{bodoprep}.

The program's output for the eigenstates is not included in the code, but presented in (\ref{allnn5}), (\ref{allee5}), (\ref{allup}) and (\ref{alldown}). 

\noindent\(
\pmb{}\\
\pmb{\text{s10}\text{:=}\{-1,-1,-1\}\left/\sqrt{3}\right.}\\
\pmb{\text{del1u}\text{:=}\{\text{d1x},\text{d1y},\text{d1z}\}*\text{ef}}\\
\pmb{\text{del2u}\text{:=}\{\text{d2x},-\text{d2y},-\text{d2z}\}*\text{ef}}\\
\pmb{\text{del3u}\text{:=}\{-\text{d3x},\text{d3y},-\text{d3z}\}*\text{ef}}\\
\pmb{\text{del4u}\text{:=}\{-\text{d4x},-\text{d4y},\text{d4z}\}*\text{ef}}\\
\pmb{}\\
\pmb{\text{t10}\text{:=}+\text{s10}}\\
\pmb{\text{eel1u}\text{:=}\{\text{e1x},\text{e1y},\text{e1z}\}*\text{ef}}\\
\pmb{\text{eel2u}\text{:=}\{\text{e2x},-\text{e2y},-\text{e2z}\}*\text{ef}}\\
\pmb{\text{eel3u}\text{:=}\{-\text{e3x},\text{e3y},-\text{e3z}\}*\text{ef}}\\
\pmb{\text{eel4u}\text{:=}\{-\text{e4x},-\text{e4y},\text{e4z}\}*\text{ef}}\\
\pmb{}\\
\pmb{\text{dd1}\text{:=}\text{del2u}+\text{del3u}+\text{del4u}-3*\text{del1u}}\\
\pmb{\text{dd2}\text{:=}\text{del1u}+\text{del3u}+\text{del4u}-3*\text{del2u}}\\
\pmb{\text{dd3}\text{:=}\text{del1u}+\text{del2u}+\text{del4u}-3*\text{del3u}}\\
\pmb{\text{dd4}\text{:=}\text{del1u}+\text{del2u}+\text{del3u}-3*\text{del4u}}\\
\pmb{}\\
\pmb{\text{ed1}\text{:=}\text{eel2u}+\text{eel3u}+\text{eel4u}-3*\text{del1u}}\\
\pmb{\text{ed2}\text{:=}\text{eel1u}+\text{eel3u}+\text{eel4u}-3*\text{del2u}}\\
\pmb{\text{ed3}\text{:=}\text{eel1u}+\text{eel2u}+\text{eel4u}-3*\text{del3u}}\\
\pmb{\text{ed4}\text{:=}\text{eel1u}+\text{eel2u}+\text{eel3u}-3*\text{del4u}}\\
\pmb{}\\
\pmb{\text{de1}\text{:=}\text{del2u}+\text{del3u}+\text{del4u}-3*\text{eel1u}}\\
\pmb{\text{de2}\text{:=}\text{del1u}+\text{del3u}+\text{del4u}-3*\text{eel2u}}\\
\pmb{\text{de3}\text{:=}\text{del1u}+\text{del2u}+\text{del4u}-3*\text{eel3u}}\\
\pmb{\text{de4}\text{:=}\text{del1u}+\text{del2u}+\text{del3u}-3*\text{eel4u}}\\
\pmb{}\\
\pmb{\text{ee1}\text{:=}\text{eel2u}+\text{eel3u}+\text{eel4u}-3*\text{eel1u}}\\
\pmb{\text{ee2}\text{:=}\text{eel1u}+\text{eel3u}+\text{eel4u}-3*\text{eel2u}}\\
\pmb{\text{ee3}\text{:=}\text{eel1u}+\text{eel2u}+\text{eel4u}-3*\text{eel3u}}\\
\pmb{\text{ee4}\text{:=}\text{eel1u}+\text{eel2u}+\text{eel3u}-3*\text{eel4u}}\\
\pmb{}\\
\pmb{\text{vdd1}\text{:=}-2*\text{dd1}+2*\text{dd1}.\text{s10}*\text{s10}}\\
\pmb{\text{vdd2}\text{:=}-2*\text{dd2}+2*\text{dd2}.\text{s10}*\text{s10}}\\
\pmb{\text{vdd3}\text{:=}-2*\text{dd3}+2*\text{dd3}.\text{s10}*\text{s10}}\\
\pmb{\text{vdd4}\text{:=}-2*\text{dd4}+2*\text{dd4}.\text{s10}*\text{s10}}\\
\pmb{}\\
\pmb{\text{ved1}\text{:=}-2*\text{ed1}+2*\text{ed1}.\text{s10}*\text{s10}}\\
\pmb{\text{ved2}\text{:=}-2*\text{ed2}+2*\text{ed2}.\text{s10}*\text{s10}}\\
\pmb{\text{ved3}\text{:=}-2*\text{ed3}+2*\text{ed3}.\text{s10}*\text{s10}}\\
\pmb{\text{ved4}\text{:=}-2*\text{ed4}+2*\text{ed4}.\text{s10}*\text{s10}}\\
\pmb{}\\
\pmb{\text{vde1}\text{:=}-2*\text{de1}+2*\text{de1}.\text{s10}*\text{s10}}\\
\pmb{\text{vde2}\text{:=}-2*\text{de2}+2*\text{de2}.\text{s10}*\text{s10}}\\
\pmb{\text{vde3}\text{:=}-2*\text{de3}+2*\text{de3}.\text{s10}*\text{s10}}\\
\pmb{\text{vde4}\text{:=}-2*\text{de4}+2*\text{de4}.\text{s10}*\text{s10}}\\
\pmb{}\\
\pmb{\text{vee1}\text{:=}-2*\text{ee1}+2*\text{ee1}.\text{s10}*\text{s10}}\\
\pmb{\text{vee2}\text{:=}-2*\text{ee2}+2*\text{ee2}.\text{s10}*\text{s10}}\\
\pmb{\text{vee3}\text{:=}-2*\text{ee3}+2*\text{ee3}.\text{s10}*\text{s10}}\\
\pmb{\text{vee4}\text{:=}-2*\text{ee4}+2*\text{ee4}.\text{s10}*\text{s10}}\\
\pmb{}\\
\pmb{\text{ss}\text{:=}-10.70000000000000000}\\
\pmb{\text{st}\text{:=}-0.07700000000000000}\\
\pmb{\text{tt}\text{:=}-0.22000000000000000}\\
\pmb{\text{jss}\text{:=}0.32000000000000000}\\
\pmb{\text{jtt}\text{:=}0.01020000000000000}\\
\pmb{\text{jst}\text{:=}0.01750000000000000}\\
\pmb{\text{ff}\text{:=}0.00049000000000000}\\
\pmb{\text{gg}\text{:=}0.00113000000000000}\\
\pmb{\text{fg}\text{:=}0.00008500000000000}\\
\pmb{\text{ne}\text{:=}-0.00000000000103000}\\
\pmb{\text{nm}\text{:=}-0.00000000000790000}\\
\pmb{\text{nt}\text{:=}0.00000000001350000}\\
\pmb{}\\
\pmb{\text{ndd1}\text{:=}-2*\text{del1u}+2*\text{del1u}.\text{s10}*\text{s10}}\\
\pmb{\text{ndd2}\text{:=}-2*\text{del2u}+2*\text{del2u}.\text{s10}*\text{s10}}\\
\pmb{\text{ndd3}\text{:=}-2*\text{del3u}+2*\text{del3u}.\text{s10}*\text{s10}}\\
\pmb{\text{ndd4}\text{:=}-2*\text{del4u}+2*\text{del4u}.\text{s10}*\text{s10}}\\
\pmb{}\\
\pmb{\text{nee1}\text{:=}-2*\text{eel1u}+2*\text{eel1u}.\text{s10}*\text{s10}}\\
\pmb{\text{nee2}\text{:=}-2*\text{eel2u}+2*\text{eel2u}.\text{s10}*\text{s10}}\\
\pmb{\text{nee3}\text{:=}-2*\text{eel3u}+2*\text{eel3u}.\text{s10}*\text{s10}}\\
\pmb{\text{nee4}\text{:=}-2*\text{eel4u}+2*\text{eel4u}.\text{s10}*\text{s10}}\\
\pmb{}\\
\pmb{\text{zx1}\text{:=} }\\
\pmb{\text{Coefficient}[\text{ss}*(2*\text{Cross}[\text{s10},\text{dd1}]+ i *\text{vdd1})+}\\
\pmb{\text{nt}*(2*\text{Cross}[\text{s10},\text{del1u}]+ i *\text{ndd1})}\\
\pmb{+\text{st}*(2*\text{Cross}[\text{s10},\text{ed1}]+ i *\text{ved1})}\\
\pmb{+\text{jss}*\text{Cross}[\text{s10},\text{dd1}]+\text{jst}*\text{Cross}[\text{s10},\text{ed1}]+\text{nm}*\text{Cross}[\text{s10},\text{del1u}]}\\
\pmb{+ i *\text{ff}*\text{dd1}+ i *\text{fg}*\text{ed1}+ i *\text{ne}*\text{del1u},\text{ef},1]}\\
\pmb{\text{zx2}\text{:=}}\\
\pmb{\text{Coefficient}[\text{ss}*(2*\text{Cross}[\text{s10},\text{dd2}]+ i *\text{vdd2})+}\\
\pmb{\text{nt}*(2*\text{Cross}[\text{s10},\text{del2u}]+ i *\text{ndd2})}\\
\pmb{+\text{st}*(2*\text{Cross}[\text{s10},\text{ed2}]+ i *\text{ved2})}\\
\pmb{+\text{jss}*\text{Cross}[\text{s10},\text{dd2}]+\text{jst}*\text{Cross}[\text{s10},\text{ed2}]+\text{nm}*\text{Cross}[\text{s10},\text{del2u}]}\\
\pmb{+ i *\text{ff}*\text{dd2}+ i *\text{fg}*\text{ed2}+ i *\text{ne}*\text{del2u},\text{ef},1]}\\
\pmb{\text{zx3}\text{:=}}\\
\pmb{\text{Coefficient}[\text{ss}*(2*\text{Cross}[\text{s10},\text{dd3}]+ i *\text{vdd3})+}\\
\pmb{\text{nt}*(2*\text{Cross}[\text{s10},\text{del3u}]+ i *\text{ndd3})}\\
\pmb{+\text{st}*(2*\text{Cross}[\text{s10},\text{ed3}]+ i *\text{ved3})}\\
\pmb{+\text{jss}*\text{Cross}[\text{s10},\text{dd3}]+\text{jst}*\text{Cross}[\text{s10},\text{ed3}]+\text{nm}*\text{Cross}[\text{s10},\text{del3u}]}\\
\pmb{+ i *\text{ff}*\text{dd3}+ i *\text{fg}*\text{ed3}+ i *\text{ne}*\text{del3u},\text{ef},1]}\\
\pmb{\text{zx4}\text{:=}}\\
\pmb{\text{Coefficient}[\text{ss}*(2*\text{Cross}[\text{s10},\text{dd4}]+ i *\text{vdd4})+}\\
\pmb{\text{nt}*(2*\text{Cross}[\text{s10},\text{del4u}]+ i *\text{ndd4})}\\
\pmb{+\text{st}*(2*\text{Cross}[\text{s10},\text{ed4}]+ i *\text{ved4})}\\
\pmb{+\text{jss}*\text{Cross}[\text{s10},\text{dd4}]+\text{jst}*\text{Cross}[\text{s10},\text{ed4}]+\text{nm}*\text{Cross}[\text{s10},\text{del4u}]}\\
\pmb{+ i *\text{ff}*\text{dd4}+ i *\text{fg}*\text{ed4}+ i *\text{ne}*\text{del4u},\text{ef},1]}\\
\pmb{}\\
\pmb{\text{zx5}\text{:=}\text{Coefficient}[\text{st}*(2*\text{Cross}[\text{s10},\text{de1}]+ i *\text{vde1})}\\
\pmb{+\text{tt}*(2*\text{Cross}[\text{s10},\text{ee1}]+ i *\text{vee1})+\text{nt}*(2*\text{Cross}[\text{s10},\text{eel1u}]+ i *\text{nee1})}\\
\pmb{+\text{jst}*\text{Cross}[\text{s10},\text{de1}]+\text{jtt}*\text{Cross}[\text{s10},\text{ee1}]+\text{nm}*\text{Cross}[\text{s10},\text{eel1u}]}\\
\pmb{+ i *\text{gg}*\text{ee1}+ i *\text{fg}*\text{de1}+ i *\text{ne}*\text{eel1u},\text{ef},1]}\\
\pmb{\text{zx6}\text{:=}\text{Coefficient}[\text{st}*(2*\text{Cross}[\text{s10},\text{de2}]+ i *\text{vde2})}\\
\pmb{+\text{tt}*(2*\text{Cross}[\text{s10},\text{ee2}]+ i *\text{vee2})+\text{nt}*(2*\text{Cross}[\text{s10},\text{eel2u}]+ i *\text{nee2})}\\
\pmb{+\text{jst}*\text{Cross}[\text{s10},\text{de2}]+\text{jtt}*\text{Cross}[\text{s10},\text{ee2}]+\text{nm}*\text{Cross}[\text{s10},\text{eel2u}]}\\
\pmb{+ i *\text{gg}*\text{ee2}+ i *\text{fg}*\text{de2}+ i *\text{ne}*\text{eel2u},\text{ef},1]}\\
\pmb{\text{zx7}\text{:=}\text{Coefficient}[\text{st}*(2*\text{Cross}[\text{s10},\text{de3}]+ i *\text{vde3})}\\
\pmb{+\text{tt}*(2*\text{Cross}[\text{s10},\text{ee3}]+ i *\text{vee3})+\text{nt}*(2*\text{Cross}[\text{s10},\text{eel3u}]+ i *\text{nee3})}\\
\pmb{+\text{jst}*\text{Cross}[\text{s10},\text{de3}]+\text{jtt}*\text{Cross}[\text{s10},\text{ee3}]+\text{nm}*\text{Cross}[\text{s10},\text{eel3u}]}\\
\pmb{+ i *\text{gg}*\text{ee3}+ i *\text{fg}*\text{de3}+ i *\text{ne}*\text{eel3u},\text{ef},1]}\\
\pmb{\text{zx8}\text{:=}\text{Coefficient}[\text{st}*(2*\text{Cross}[\text{s10},\text{de4}]+ i *\text{vde4})}\\
\pmb{+\text{tt}*(2*\text{Cross}[\text{s10},\text{ee4}]+ i *\text{vee4})+\text{nt}*(2*\text{Cross}[\text{s10},\text{eel4u}]+ i *\text{nee4})}\\
\pmb{+\text{jst}*\text{Cross}[\text{s10},\text{de4}]+\text{jtt}*\text{Cross}[\text{s10},\text{ee4}]+\text{nm}*\text{Cross}[\text{s10},\text{eel4u}]}\\
\pmb{+ i *\text{gg}*\text{ee4}+ i *\text{fg}*\text{de4}+ i *\text{ne}*\text{eel4u},\text{ef},1]}\\
\pmb{}\\
\pmb{\text{S535}\text{:=}\text{Flatten}[i\{\text{zx1},\text{zx2},\text{zx3},\text{zx4},\text{zx5},\text{zx6},\text{zx7},\text{zx8}\}]}\\
\pmb{}\\
\pmb{\text{Eigensystem}[}\\
\pmb{\{}\\
\pmb{\text{Coefficient}[\text{S535},\text{d1x},1],}\\
\pmb{\text{Coefficient}[\text{S535},\text{d1y},1],}\\
\pmb{\text{Coefficient}[\text{S535},\text{d1z},1],}\\
\pmb{\text{Coefficient}[\text{S535},\text{d2x},1],}\\
\pmb{-\text{Coefficient}[\text{S535},\text{d2y},1],}\\
\pmb{-\text{Coefficient}[\text{S535},\text{d2z},1],}\\
\pmb{-\text{Coefficient}[\text{S535},\text{d3x},1],}\\
\pmb{\text{Coefficient}[\text{S535},\text{d3y},1],}\\
\pmb{-\text{Coefficient}[\text{S535},\text{d3z},1],}\\
\pmb{-\text{Coefficient}[\text{S535},\text{d4x},1],}\\
\pmb{-\text{Coefficient}[\text{S535},\text{d4y},1],}\\
\pmb{\text{Coefficient}[\text{S535},\text{d4z},1],}\\
\pmb{\text{Coefficient}[\text{S535},\text{e1x},1],}\\
\pmb{\text{Coefficient}[\text{S535},\text{e1y},1],}\\
\pmb{\text{Coefficient}[\text{S535},\text{e1z},1],}\\
\pmb{\text{Coefficient}[\text{S535},\text{e2x},1],}\\
\pmb{-\text{Coefficient}[\text{S535},\text{e2y},1],}\\
\pmb{-\text{Coefficient}[\text{S535},\text{e2z},1],}\\
\pmb{-\text{Coefficient}[\text{S535},\text{e3x},1],}\\
\pmb{\text{Coefficient}[\text{S535},\text{e3y},1],}\\
\pmb{-\text{Coefficient}[\text{S535},\text{e3z},1],}\\
\pmb{-\text{Coefficient}[\text{S535},\text{e4x},1],}\\
\pmb{-\text{Coefficient}[\text{S535},\text{e4y},1],}\\
\pmb{\text{Coefficient}[\text{S535},\text{e4z},1]}\\
\pmb{\}}\\
\pmb{]}\)

\noindent\(\pmb{\{170.794,170.794,170.794,4.35497,4.35497,4.35497,1.74351,}\\
\pmb{1.33497,1.33497,1.33497,0.10551,0.097825,0.097825,0.097825,}\\
\pmb{0.00477782,0.00477782,0.00477782,0.00221218,0.00221218,}\\
\pmb{0.00221218,0.00051,4.7123*10{}^{\wedge}-11,8.92766*10{}^{\wedge}-12,1.02624*10{}^{\wedge}-12\}}\)

\section{Supplementary material II: Application to CKM and PMNS matrices} 

In discussions of neutrino masses there is always the question whether they are of Dirac or Majorana type. Within the tetron model, neutrinos have the same spacetime properties as the other quarks and leptons, because all isospin excitations inherit their SO(3,1) transformation properties from the underlying octonion representation of SO(6,1) - which is Dirac.

As well known there is a mixing between the flavor and mass eigenstates of the 3 neutrino species, and this can be described by a unitary matrix, the PMNS neutrino mixing matrix\cite{maki,giganti}. The experimentally relevant quantities are the absolute values of the matrix elements, which describe the amount of admixture of the flavor into mass eigenstates, and the leptonic Jarlskog invariant which describes any possible CP violation in the leptonic sector.

Unfortunately, within the SM the values of the mixing parameters cannot be predicted. 

\vspace{0.2cm}
{\bf Leading symmetric approximation for the PMNS matrix}

In a first step a leading order result for the mixing matrix will be derived which is
\begin{eqnarray} 
V_{PMNS}&=&\exp \Biggl\{\frac{i}{{\sqrt{3}}} \begin{bmatrix}
   0 & 1 & 0  \\
    1 & 1 & -1  \\
     0 &  -1 & -1  \\
\end{bmatrix}\Biggr\} \nonumber \\
&=& \begin{bmatrix}
   0.8467-i 0.0300 & -0.1489+i 0.4861 & 0.1532-i 0.00051  \\
    -0.1489-i 0.4861 & 0.5446+i 0.4568 & -0.00433 - i 0.4858  \\
      0.1532-i 0.00051 &  -0.00433 - i 0.4858 & 0.6892-i 0.5153  \\
\end{bmatrix}
\label{pmn0a}
\end{eqnarray}
while an improved formula will be given later in (\ref{p0imp}). 

The leading order expression (\ref{pmn0a}) is a complex, symmetric and unitary matrix, and the absolute values of the matrix elements can be calculated numerically and compared to measurements\cite{pdg} 
\begin{eqnarray} 
\begin{bmatrix}
   0.843 & 0.510 & 0.153  \\
    0.510 & 0.711 & 0.486  \\
      0.153 &  0.486 & 0.861  \\
\end{bmatrix}
\quad vs. \quad
\begin{bmatrix}
   0.80-0.85 & 0.51-0.58 & 0.142-0.155  \\
    0.23-0.51 & 0.46-0.69 & 0.63-0.78  \\
      0.25-0.53 &  0.47-0.70 & 0.61-0.76  \\
\end{bmatrix}
\label{pmn1a}
\end{eqnarray} 
By inspection one concludes that the agreement is reasonable but not optimal, with the 23 entry being the most critical. 
The first row, which is best measured, is also best fitting. Concerning the other rows, the experimental results in (\ref{pmn1a}) are non-symmetric, though with very large errors. It will be described later, in connection with (\ref{p0imp}) and (\ref{pmnxxa}), how (\ref{pmn0a}) can be improved by additional non-symmetric contributions so that complete agreement within the errors is obtained.

Besides the absolute values, which describe the amount of admixture of the flavor into mass eigenstates, the only other experimentally relevant quantity of the PMNS matrix is the leptonic Jarlskog invariant\cite{jarls} which describes any possible CP violation in the leptonic sector. A prediction for $J_{PMNS}$ can be calculated from  (\ref{pmn0a}) as  
\begin{eqnarray} 
J_{PMNS}=\Im (V_{e1} V_{\mu 2} \bar V_{e2} \bar V_{\mu 1} ) =-0.0106  
\label{pmn2a}
\end{eqnarray} 
This value is large as compared to the Jarlskog parameter of the CKM matrix\cite{pdg}. $J_{PMNS}$ has not been measured so far, although there are experimental indications that leptonic CP violation is indeed rather large\cite{jarl2}. 

\vspace{0.2cm}
{\bf Motivation and proof}

As explained before, quark and lepton masses and mass eigenstates are obtained by diagonalizing the 24 equations $d \vec \delta/dt$ for the isospin excitations $\vec \delta=\vec Q -\langle \vec Q \rangle$. While the masses correspond to the eigenvalues, CKM and PMNS mixings can be deduced from the eigenvectors. The relations between the excitations $\vec \delta$, the mass eigenstates and the weak interaction eigenstates are clarified in the following discussion. Thereby, the result (\ref{pmn0a}) for the PMNS matrix will be obtained. 

The first step is to label the quark and lepton mass states in terms of the vectors $\vec \delta$. More in detail, the following definitions are used: 
\begin{eqnarray} 
\ket{\vec S}= \vec \delta _L  \qquad\qquad\qquad \ket{\vec T}=  \vec \delta _R 
\label{nn77}
\end{eqnarray}
Dirac's notation with bra and ket states is applied here to make the mixing relations more transparent. In fact, (\ref{nn77}) are orthonormal vector states and can be used to write down the equations for the neutrino mass eigenstates, as obtained from the diagonalization procedure
\begin{eqnarray} 
\ket{\nu_{e,m}}&=&\frac{1}{\sqrt{6}} [(\ket{S_x} +\ket{T_x}) +  (\ket{S_y} + \ket{T_y})  + (\ket{S_z} + \ket{T_z}) ] \nonumber \\
\ket{\nu_{\mu, m}}&=&\frac{1}{\sqrt{6}} [ (\ket{S_x} + \ket{T_x}) +\omega (\ket{S_y} + \ket{T_y}) + \bar \omega (\ket{S_z} + \ket{T_z})] \nonumber \\
\ket{\nu_{\tau, m}}&=& \frac{1}{\sqrt{6}} [(\ket{S_x} +\ket{T_x}) +\bar\omega (\ket{S_y} + \ket{T_y}) + \omega  (\ket{S_z} + \ket{T_z}) ] 
\label{allnn5}
\end{eqnarray}
The corresponding result for the charged leptons is
\begin{eqnarray} 
\ket{e_m}&=&\frac{1}{\sqrt{6}} [  (\ket{T_x}-\ket{S_x}) +  (\ket{T_y}-\ket{S_y})  + (\ket{T_z}-\ket{S_z}) ] \nonumber \\
\ket{\mu_m}&=& \frac{1}{\sqrt{6}}[(\ket{T_x}-\ket{S_x}) + \omega (\ket{T_y}-\ket{S_y})  + \bar \omega (\ket{T_z}-\ket{S_z}) ] \nonumber  \\
\ket{\tau_m}&=& \frac{1}{\sqrt{6}}[(\ket{T_x}-\ket{S_x}) + \bar \omega (\ket{T_y}-\ket{S_y})  +\omega (\ket{T_z}-\ket{S_z}) ] 
\label{allee5}
\end{eqnarray}

The appearance of the complex numbers
\begin{eqnarray} 
\omega=-\frac{1-i\sqrt{3}}{2} \qquad\qquad\qquad \bar\omega=-\frac{1+i\sqrt{3}}{2}
\label{mack151}
\end{eqnarray}
corresponding to rotations by 120 and 240 degrees are an effect of the underlying tetrahedral symmetry. They turn the expressions (\ref{allnn5}) and (\ref{allee5}) into symmetry adapted functions.

The lepton mass states actually can be brought to the much more compact form 
\begin{eqnarray} 
\begin{bmatrix}  \ket{\nu_{e m}}\\     \ket{\nu_{\mu m} }   \\  \ket{\nu_{\tau m} }    \\ \end{bmatrix}
=Z\begin{bmatrix}  \ket{V_x}\\     \ket{V_y}   \\     \ket{V_z}    \\ \end{bmatrix}
\qquad\qquad\qquad\qquad
\begin{bmatrix}  \ket{e_m}\\     \ket{\mu_m}   \\     \ket{\tau_m}    \\ \end{bmatrix}
=Z\begin{bmatrix}  \ket{A_x}\\     \ket{A_y}   \\     \ket{A_z}    \\ \end{bmatrix}
\label{fo77}
\end{eqnarray}
by using the quantities 
\begin{eqnarray} 
\ket{\vec V}= \frac{1}{\sqrt{2}}(\ket{\vec S}+\ket{\vec T})  \qquad \qquad\qquad
\ket{\vec A}= \frac{1}{\sqrt{2}}(\ket{\vec T}-\ket{\vec S}) 
\label{nn77a}
\end{eqnarray}
and the $Z_3$ Fourier transform matrices
\begin{eqnarray} 
Z=\frac{1}{\sqrt{3}}
\begin{bmatrix}
   1 & 1 & 1  \\
    1 & \omega & \bar\omega   \\
    1 &  \bar\omega &  \omega   \\
\end{bmatrix}
\qquad \qquad 
Z^\dagger=\frac{1}{\sqrt{3}}
\begin{bmatrix}
   1 & 1 & 1  \\
  1 &  \bar\omega & \omega   \\
   1 &   \omega &  \bar\omega   \\
\end{bmatrix}
\label{matz35}
\end{eqnarray}
It is interesting to note that the eigenfunctions (\ref{allnn5}), (\ref{allee5}) and (\ref{fo77}) are stable against variations of all the isospin couplings one may use in the Hamiltonian H in (\ref{txxm32}). In consequence, the neutrino mixing matrix does not depend on any fermion mass values. This implies a stable and unambiguous prediction for the PMNS matrix and is in contrast to the CKM matrix in the quark sector, where a mass dependence shows up. 

As well known, the defining equation for the PMNS matrix is
\begin{eqnarray} 
\begin{bmatrix}  \bra{ \nu_{ew}} & \bra{ \nu_{\mu w}} & \bra{ \nu_{\tau w}} \\ \end{bmatrix} W_\mu^+ 
\begin{bmatrix}  \ket{e_w} \\ \ket{\mu_w}  \\ \ket{\tau_w} \\ \end{bmatrix}
=\begin{bmatrix}  \bra{ \nu_{em}} & \bra{ \nu_{\mu m}} & \bra{ \nu_{\tau m}} \\ \end{bmatrix} W_\mu^+ V_{PMNS}
\begin{bmatrix}  \ket{e_m} \\ \ket{\mu_m}  \\ \ket{\tau_m} \\ \end{bmatrix}
\label{wbwwc}
\end{eqnarray} 
where the index $w$ denotes weak interaction eigenstates, and it is understood that we talk about left handed fields only. The mixing matrix is formally given by
\begin{eqnarray} 
V_{PMNS}=V_N V_L^\dagger = 
\begin{bmatrix}
V_{1e} & V_{1\mu} & V_{1\tau} \\
V_{2e} & V_{2\mu} & V_{2\tau} \\
V_{3e} & V_{3\mu} & V_{3\tau}
\end{bmatrix} 
\label{wb111c}
\end{eqnarray} 
where
\begin{eqnarray} 
V_N=
\begin{bmatrix}
\bra{\nu_{e m}}  \\ \bra{\nu_{\mu m}}\\ \bra{\nu_{\tau m}} \\ 
\end{bmatrix}
\begin{bmatrix}
\ket{\nu_{e w}}  & \ket{\nu_{\mu w}}& \ket{\nu_{\tau w}} \\ 
\end{bmatrix}
\qquad
V_L^\dagger=
\begin{bmatrix}
\bra{e_w}  \\ \bra{\mu_w} \\ \bra{\tau_w} \\ 
\end{bmatrix}
\begin{bmatrix}
\ket{e_m}  & \ket{\mu_m}& \ket{\tau_m} \\ 
\end{bmatrix}
\label{wb222c}
\end{eqnarray} 
Replacing the mass eigenstates by the isospin excitations according to (\ref{fo77}) one obtains
\begin{eqnarray} 
V_{PMNS}=Z 
\Biggl\{
\begin{bmatrix}
\bra{V_x}  \\ \bra{V_y} \\ \bra{V_z} \\ 
\end{bmatrix}
\begin{bmatrix}  \ket{ \nu_{ew}} & \ket{ \nu_{\mu w}} & \ket{ \nu_{\tau w}} \\ \end{bmatrix}
\begin{bmatrix}
\bra{e_w}  \\ \bra{\mu_w} \\ \bra{\tau_w} \\ 
\end{bmatrix}
\begin{bmatrix}
\ket{A_x}  & \ket{A_y}& \ket{A_z} \\ 
\end{bmatrix}
\Biggr\} Z^\dagger
\label{wb333c}
\end{eqnarray} 
By inspection one sees that (\ref{wb333c}) exactly compensates all the matrix transformations in (\ref{wbwwc}) and (\ref{fo77}) so as to maintain lepton universality and keep the weak current diagonal in the weak eigenstates.

The brace in (\ref{wb333c}) comprises a matrix of expectation values of the form \begin{eqnarray} 
Y:=\begin{bmatrix} \bra{V_x}  \\ \bra{V_y} \\ \bra{V_z} \\ \end{bmatrix}
\mathcal{O}
\begin{bmatrix}
\ket{A_x}  & \ket{A_y}& \ket{A_z} \\ 
\end{bmatrix}
\label{wb8c}
\end{eqnarray} 
where the inner product 
\begin{eqnarray} 
\mathcal{O}:= 
\begin{bmatrix}  \ket{ \nu_{ew}} & \ket{ \nu_{\mu w}} & \ket{ \nu_{\tau w}} \\ \end{bmatrix}
\begin{bmatrix}
\bra{e_w}  \\ \bra{\mu_w} \\ \bra{\tau_w} \\ 
\end{bmatrix}
\label{dyop1}
\end{eqnarray} 
is a dyadic 1-dimensional operator which acts between the complex 3-dimensional spaces of charged lepton ($\sim \vec S -\vec T$) and antineutrino ($\sim \vec S +\vec T$) states. One may say that it contains all information about what the charged W-boson does to the lepton fields: it changes isospin, mixes families and so on. Weak SU(2) and tetrahedral symmetry force $\mathcal{O}$ to have the form
\begin{eqnarray} 
\mathcal{O}&=&\ket { S_x} \bra  { T_x} + \ket { S_y} \bra  { T_y} + \ket { S_z} \bra  { T_z} -  \ket { T_x} \bra  { S_x} - \ket { T_y} \bra  { S_y} - \ket { T_z} \bra  { S_z}\nonumber \\
&& +\frac{i}{\sqrt{3}}[\ket { S_y} \bra  { S_z} + \ket { S_z} \bra  { S_y} - \ket { T_y} \bra  { T_z} - \ket { T_z} \bra  { T_y}]\nonumber \\
&& +\frac{i}{\sqrt{3}}[\omega \ket { S_x} \bra  { S_y} + \bar \omega \ket { S_y} \bra  { S_x} - \omega \ket { T_x} \bra  { T_y} -  \bar \omega \ket { T_y} \bra  { T_x}]\nonumber \\
&& +\frac{i}{\sqrt{3}}[ \bar \omega \ket { S_x} \bra  { S_z} + \omega \ket { S_z} \bra  { S_x} -  \bar \omega \ket { T_x} \bra  { T_z} - \omega \ket { T_z} \bra  { T_x}]  
\label{wb1}
\end{eqnarray} 
In order to derive (\ref{wb1}) one has to note that SU(2) invariance allows the appearance of dot products and triple products only. The coefficients of these products are then dictated by the tetrahedral symmetry of the isospin vectors. For example, to derive the triple product coefficients one should remember that the $W^+$-boson is defined in the 3 internal dimensions in an analogous manner as a plus circularly polarized wave in 3 spatial dimensions, namely by means of an (internal) `polarization vector' $\vec e_+=(\vec e_1 + i \vec e_2)/\sqrt{2}$ which is perpendicular to the axis of quantization, in this case given by $\sim (1,1,1)$.
\begin{eqnarray} 
\vec e_1 = \frac{1}{\sqrt{2}} (0,1,-1) \qquad  \qquad \vec e_2=\frac{1}{\sqrt{6}} (-2,1,1)
\label{pol11}
\end{eqnarray} 
Introducing the vector
\begin{eqnarray} 
\vec \Omega = \frac{1}{\sqrt{3}} (1,\omega,\bar \omega) 
\label{pol11999}
\end{eqnarray} 
allowed contributions to $\mathcal{O}$ are of the triple product form
\begin{eqnarray} 
&& \varepsilon_{ijk}  \frac{1}{\sqrt{2}}(\vec e_1+i \vec e_2)_i \ket { Q_j} \bra  { Q'_k}
= -\frac{i}{\sqrt{3}} \, \vec \Omega (\vec Q \times \vec Q') 
= -\frac{i}{\sqrt{3}} \, [\,\ket { Q'_y} \bra  { Q_z} - \ket { Q'_z} \bra  { Q_y}\nonumber \\
&& \qquad\qquad -\omega (\ket { Q'_x} \bra  { Q_z} - \ket { Q'_z} \bra  { Q_x})
+ \bar \omega (\ket { Q'_x} \bra  { Q_y} - \ket { Q'_y} \bra  { Q_x}\,) \,]
\label{pol12}
\end{eqnarray} 
for the ket and bra states belonging to any 2 internal angular momenta $Q$ and $Q'$. These contributions are anti-hermitian, and care must be taken in the definition of the complex triple product when using complex conjugation in the determination of $\mathcal{O}$. 

Note that $\mathcal{O}$ as given in (\ref{wb1}) is universal in the sense that it depends only on properties of the $\Psi$ field, and therefore will appear in identical form within the quark sector and the calculation of the CKM matrix. This fact reflects the quark lepton universality of the W-boson interactions.

Inserting (\ref{wb1}) into (\ref{wb8c}) one obtains 
\begin{eqnarray} 
Y=\begin{bmatrix} \bra{V_x}  \\ \bra{V_y} \\ \bra{V_z} \\ \end{bmatrix}
\mathcal{O}
\begin{bmatrix}
\ket{A_x}  & \ket{A_y}& \ket{A_z} \\ 
\end{bmatrix}
=I+X
\label{wb8cc}
\end{eqnarray} 
i.e. a sum of a hermitian part (the unit matrix $I$) and an anti-hermitian matrix
\begin{eqnarray} 
X=-\frac{i}{\sqrt{3}}\begin{bmatrix}
   0 & \bar\omega & \omega  \\
    \omega & 0 & 1  \\
     \bar \omega &  1 & 0  \\
\end{bmatrix} 
\label{matz325}
\end{eqnarray}
The invariant structure which gives the unit matrix in (\ref{wb8cc}) is the dot product, while the invariant structure belonging to the anti-hermitian contribution X is the triple product. The unit matrix corresponds to no mixing at all, so the origin of a non-trivial PMNS matrix is to be found solely in the triple product terms (\ref{pol12}). 

Since the result (\ref{wb8cc}) is not unitary but anti-hermitian, an exponentiation suggests itself which gives a unitary PMNS matrix of the form
\begin{eqnarray} 
V_{PMNS}&=&Z e^{X} Z^\dagger =e^{Z X Z^\dagger}  \nonumber \\
&=& \frac{1}{3}\begin{bmatrix}
   1 & 1 & 1  \\
    1 & \omega & \bar\omega   \\
    1 &  \bar\omega &  \omega   \\
    \end{bmatrix}
\exp \Biggl\{\frac{-i}{{\sqrt{3}}} \begin{bmatrix}
   0 & \bar\omega & \omega  \\
    \omega & 0 & 1  \\
     \bar \omega &  1 & 0  \\
\end{bmatrix}\Biggr\}
\begin{bmatrix}
   1 & 1 & 1  \\
  1 &  \bar\omega & \omega   \\
   1 &   \omega &  \bar\omega   \\
\end{bmatrix}  \nonumber \\
&=& \begin{bmatrix}
   0.8467-i 0.0300 & -0.1489+i 0.4861 & 0.1532-i 0.00051  \\
    -0.1489-i 0.4861 & 0.5446+i 0.4568 & -0.00433 - i 0.4858  \\
      0.1532-i 0.00051 &  -0.00433 - i 0.4858 & 0.6892-i 0.5153  \\
\end{bmatrix}
\label{pmn00a}
\end{eqnarray}
identical to what was claimed in (\ref{pmn0a}).


\vspace{0.2cm}
{\bf Improved formula for the PMNS matrix}

So far only dot product and triple product terms (\ref{pol12}) have been considered as contributing to the operator (\ref{wb1}) and the PMNS result. Actually, there is a third kind of term that needs consideration. Using $\vec\Omega^2 =0$ and the same normalization as in (\ref{pol12}) it is of the form 
\begin{eqnarray} 
-(\vec \Omega \times \vec Q) \, (\vec \Omega \times \vec Q')= (\vec \Omega \vec Q) \, (\vec \Omega \vec Q')
\label{w8899}
\end{eqnarray} 
In the microscopic model, quark and lepton masses are related to torsional, Heisenberg and Dzyaloshinskii isospin interactions of the fundamental $\Psi$ field. Furthermore, as shown in \cite{bodoprep}, these three types of interactions completely fix the structure of the model. 

This fact is reflected in the contributions to the operator $\mathcal{O}$: while the dot products and triple products appearing in (\ref{wb1}) parallel the torsional and Heisenberg interactions, (\ref{w8899}) corresponds to the Dzyaloshinskii Hamiltonian. Working out the products $\ket {Q_i} \bra  {Q'_j}$ arising from (\ref{w8899}), it leads to an additional contribution to (\ref{wb1}) which can be comprised by a matrix
\begin{eqnarray} 
D:= \frac{1}{3}
 \begin{bmatrix}
   1 & \omega & \bar \omega  \\
    \omega & \bar \omega & 1  \\
     \bar \omega & 1 & \omega  \\
\end{bmatrix}
\label{ddma1}
\end{eqnarray}
The role of D for (\ref{w8899}) is analogous to that of X for the triple product term.
Combining the X and D contributions an improved formula for the PMNS matrix is obtained
\begin{eqnarray} 
V_{PMNS}=
\exp \Biggl\{  \frac{1}{3}\begin{bmatrix}
   0 & 0 & 0  \\
    0 & 0 & 1  \\
     0 &  -1 & 0  \\
\end{bmatrix}\Biggr\} 
\,
\exp \Biggl\{\frac{i}{{\sqrt{3}}} \begin{bmatrix}
   0 & 1 & 0  \\
    1 & 1 & -1  \\
     0 &  -1 & -1  \\
\end{bmatrix}\Biggr\} 
\label{p0imp}
\end{eqnarray}
This represents a complex and unitary matrix whose absolute value matrix $|V_{PMNS}|$ is not symmetric, in contrast to (\ref{pmn0a}). Its elements are given by
\begin{eqnarray} 
\begin{bmatrix}
   0.847 & 0.510 & 0.153  \\
    0.468 & 0.581 & 0.666  \\
      0.251 &  0.635 & 0.730  \\
\end{bmatrix}
\quad vs. \quad
\begin{bmatrix}
   0.80-0.85 & 0.51-0.58 & 0.142-0.155  \\
    0.23-0.51 & 0.46-0.69 & 0.63-0.78  \\
      0.25-0.53 &  0.47-0.70 & 0.61-0.76  \\
\end{bmatrix}
\label{pmnxxa}
\end{eqnarray} 
and fit the phenomenological numbers to within one standard error. 

The value of the leptonic Jarlskog invariant now is 
\begin{eqnarray} 
J_{PMNS}=0.0454
\label{pnw33}
\end{eqnarray} 
Thus, while the improvement (\ref{p0imp}) only moderately corrects  the absolute values, it strongly modifies the prediction for $J_{PMNS}$. This is because - in contrast to the absolute values - the Jarlskog invariant is dominated by higher orders of the exponentials.  

\vspace{0.2cm}
{\bf Application to the quark sector}

Mixing in the quark sector has been known since the time of Cabibbo\cite{cab}. Although the mixing percentages are smaller, it is much better measured than in the lepton sector. On the other hand, concerning theory, the predictions for the CKM mixing elements  in the present model are somewhat more difficult to obtain, though parts of the arguments for leptons can be taken over to the quark sector. The idea is again that the mixing matrix counterbalances the deviation of the mass eigenstates from the weak eigenstates in such a way that the charged current effectively acts diagonal on the isospin operators (\ref{nn77}). The main complication is the appearance of mass dependent factors in the quark eigenstates, see below. 

The CKM matrix is defined analogously to the PMNS matrix (\ref{wb111c}) and (\ref{wb222c}) 
\begin{eqnarray}
V_{CKM}&=&V_{U} V_{D}^\dagger = 
\begin{bmatrix}
V_{ud} & V_{us} & V_{ub} \\
V_{cd} & V_{cs} & V_{cb} \\
V_{td} & V_{ts} & V_{tb}
\end{bmatrix} \nonumber \\
&=& \begin{bmatrix}
\langle u_m | u_w \rangle & \langle u_m | c_w \rangle & \langle u_m | t_w \rangle\\
\langle c_m | u_w \rangle & \langle c_m | c_w \rangle & \langle c_m | t_w \rangle\\
\langle t_m | u_w \rangle & \langle t_m | c_w \rangle & \langle t_m | t_w \rangle
\end{bmatrix}
\begin{bmatrix}
\langle d_w | d_m \rangle & \langle d_w | s_m \rangle & \langle d_w | b_m \rangle\\
\langle s_w | d_m \rangle & \langle s_w | s_m \rangle & \langle s_w | b_m \rangle\\
\langle b_w | d_m \rangle & \langle b_w | s_m \rangle & \langle b_w | b_m \rangle
\end{bmatrix}
\label{mmccm}
\end{eqnarray}
where $m$ denotes mass eigenstates (the physical states) and $w$ weak interaction eigenstates. 

Solving the eigenvalue problem (\ref{txxm32}) leads to mass eigenstates for the up-type quarks
\begin{eqnarray} 
u_m&=&\frac{1}{\sqrt{3}\sqrt{1+\epsilon_{1}^2}}  [(\ket{S_x} +\epsilon_{1} \ket{T_x}) +  (\ket{S_y} + \epsilon_{1} \ket{T_y})  + (\ket{S_z} + \epsilon_{1} \ket{T_z}) ] 
\nonumber \\
c_m&=& \frac{1}{\sqrt{3}\sqrt{1+\epsilon_{2}^2}} [ (\ket{S_x} +\epsilon_2 \ket{T_x}) +\omega (\ket{S_y} + \epsilon_2 \ket{T_y}) + \bar \omega (\ket{S_z} + \epsilon_2 \ket{T_z})] \nonumber \\
t_m&=& \frac{1}{\sqrt{3}\sqrt{1+\epsilon_{3}^2}} [(\ket{S_x} +\epsilon_{3} \ket{T_x}) +\bar\omega (\ket{S_y} + \epsilon_{3} \ket{T_y}) + \omega  (\ket{S_z} + \epsilon_{3} \ket{T_z}) ]
\label{allup}
\end{eqnarray}
and for the down quarks
\begin{eqnarray} 
d_m&=&\frac{1}{\sqrt{3}\sqrt{1+\epsilon_{1}^2}} [  (\ket{T_x}-\epsilon_1 \ket{S_x}) +  (\ket{T_y}-\epsilon_1 \ket{S_y})  + (\ket{T_z}-\epsilon_1 \ket{S_z}) ] \nonumber \\
s_m&=& \frac{1}{\sqrt{3}\sqrt{1+\epsilon_{2}^2}} [(\ket{T_x}-\epsilon_2 \ket{S_x}) + \omega (\ket{T_y}-\epsilon_2 \ket{S_y})  + \bar \omega (\ket{T_z}-\epsilon_2 \ket{S_z}) ] \nonumber  \\
b_m&=& \frac{1}{\sqrt{3}\sqrt{1+\epsilon_{3}^2}} [(\ket{T_x}-\epsilon_3 \ket{S_x}) + \bar \omega (\ket{T_y}-\epsilon_3 \ket{S_y})  +\omega (\ket{T_z}-\epsilon_3 \ket{S_z}) ]
\label{alldown}
\end{eqnarray}
Three coefficients $\epsilon_{1,2,3}$ appear in these equations, which depend on the quark and even on the lepton masses. They can be calculated within the model. Namely, each $\epsilon_{i}$ to a very good approximation only depends on the quark and charged lepton masses of the i-th family. More precisely, using the symbolic version of the Mathematica program in Appendix I one can derive the formula
\begin{eqnarray} 
\epsilon_{i} = \frac{1}{6} \frac{M_{Li}}{M_{Ui}+M_{Di}}
\label{sm14exxx}
\end{eqnarray}
where $M_{Ui}$, $M_{Di}$ and $M_{Li}$ denote the corresponding masses within family i.

By inspection one sees that the lepton eigenfunctions (\ref{allnn5}) and (\ref{allee5}) are recovered from  (\ref{allup}) and (\ref{alldown}) by chosing $\epsilon_{3} = \epsilon_{2} = \epsilon_{1} = 1$. It should be stressed, however, that this is only formally true, because the quark states are defined in a different space than the lepton states. The point is that for simplicity reference has been made so far to only one of the four isospins I, II, III and IV on the tetrahedral structure. While the contributions from I-IV to the lepton states are identical and of the form I+II+III+IV, the generic form of the quark states turns out to be 3$\times$I-II-III-IV, 3$\times$II-I-III-IV and 3$\times$III-II-IV for the 3 colors, respectively.  

Knowing the eigenstates (\ref{allup}) and (\ref{alldown}) one may write down the CKM matrix in an analogous fashion as the PMNS matrix (\ref{wb333c}) for leptons
\begin{eqnarray} 
V_{CKM}=\biggl\{ R Z  
\begin{bmatrix} \bra{S_x}  \\ \bra{S_y} \\ \bra{S_z} \\ \end{bmatrix}
+REZ \begin{bmatrix} \bra{T_x}  \\ \bra{T_y} \\ \bra{T_z} \\ \end{bmatrix} 
\biggr\}
\begin{bmatrix} \ket{ u_w}  & \ket{ c_w}& \ket{ t_w} \\ \end{bmatrix}
\begin{bmatrix} \bra{d_w}  \\ \bra{s_w} \\ \bra{b_w} \\ \end{bmatrix}
\times \nonumber\\
\times \biggl\{ \begin{bmatrix} \ket{T_x}  & \ket{T_y}& \ket{T_z} \\ \end{bmatrix}
 Z^\dagger R - \begin{bmatrix} \ket{S_x}  & \ket{S_y}& \ket{S_z} \\ \end{bmatrix}
Z^\dagger E R \biggr\}
\label{wckm33}
\end{eqnarray} 
where the matrices 
\begin{eqnarray} 
E:=\begin{bmatrix}
   \epsilon_1 & 0 & 0  \\
    0 & \epsilon_2 & 0  \\
    0 &  0 & \epsilon_3  \\
\end{bmatrix}
\qquad\qquad
R:=\begin{bmatrix}
   \frac{1}{\sqrt{1+\epsilon_1^2}} & 0 & 0  \\
    0 & \frac{1}{\sqrt{1+\epsilon_2^2}} & 0  \\
    0 &  0 & \frac{1}{\sqrt{1+\epsilon_3^2}} \\
\end{bmatrix}
\label{abktrua}
\end{eqnarray} 
have been introduced.

Just as in the case of leptons (\ref{dyop1}) there is a 1-dimensional dyadic transformation
\begin{eqnarray} 
\mathcal{O} = 
\begin{bmatrix} \ket{u_w}  & \ket{c_w}& \ket{t_w} \\ \end{bmatrix}
\begin{bmatrix} \bra{d_w}  \\ \bra{s_w} \\ \bra{b_w} \\ \end{bmatrix}
\label{dyop2}
\end{eqnarray} 
which operates between the 3-dimensional spaces of up- and down-type quark states. Due to quark-lepton universality, when expressed in terms of operators $\vec S$ and $\vec T$, the operator $\mathcal{O}$ for quarks must be identical to what was used for leptons in (\ref{wb1}).

Restricting, for a moment, on the dot and triple product contributions (\ref{wb1}) as input, one may then calculate $V_{CKM}$ given in (\ref{wckm33}) to be 
\begin{eqnarray} 
V_{CKM}=I + RZX Z^\dagger E R + REZXZ^\dagger R
\rightarrow \exp \{RZX Z^\dagger E R + REZXZ^\dagger R\}
\label{wck41}
\end{eqnarray} 
where I is the 3$\times$3 unit matrix arising from the dot product terms in (\ref{wb1}). The other terms in (\ref{wck41}) are the anti-hermitian contributions from the triple product in (\ref{pol12}) and (\ref{wb1}). They replace the expression $ZXZ^\dagger$ in (\ref{pmn00a}) for leptons. 

Just as in the case of leptons one may improve on this result by including the contributions from (\ref{w8899}), in order to obtain the desired non-symmetric contributions to $|V_{CKM}|$. The improved formula for the CKM matrix reads
\begin{eqnarray} 
V_{CKM}=\exp \{ 2[RZD Z^\dagger E R - REZD^\dagger Z^\dagger R] \}
\exp \{RZX Z^\dagger E R + REZXZ^\dagger R\}
\label{wck41111}
\end{eqnarray} 
In contrast to X in (\ref{matz325}) the matrix D in (\ref{ddma1}) is not anti-hermitian. This fact has been accounted for in the first exponential factor.

Eq. (\ref{wck41111}) allows to evaluate $|V_{CKM}|$ using appropriate values for the fermion masses entering (\ref{sm14exxx}). It must be noted, however, that the low energy values of the $\epsilon_{i}$ are not useful in this context. Instead one should use running masses near the Planck scale, because the dynamics generates fermion masses originally at Planck scale distances\footnote{A GUT scale is not present in the model. There is only the Fermi scale, defined as the interaction energy of the isospin vectors, and the Planck scale, defined as the binding energy of the fields $\Psi$\cite{bodohiggs}.}. Unfortunately, the predictions for running masses are not very precise because higher order contributions become appreciable at very large scales. Nevertheless, I am using results from the literature\cite{runnmass,juarez} to determine the $\epsilon_{i}$ at high scales.
\begin{eqnarray} 
\epsilon_{1} = 0.35  \qquad \epsilon_{2} =0.070 \qquad \epsilon_{3} =0.0040
\label{sm15}
\end{eqnarray}
unfortunately with a large theoretical error, whose magnitude even is hard to estimate\cite{juarez}. The numbers are for a 2HDM (2 Higgs doublet model) which is known to be the low-energy limit of the microscopic model\cite{bodohiggs}. They exhibit a family hierarchy which will be seen to induce a corresponding hierarchy in the mixing of the quark families. Actually, as discussed in earlier work\cite{bodomasses}, this is to be expected within the present model due to the large top mass which forces the up- and down-type mass eigenstates to be approximately $\sim\vec S$ and $\sim\vec T$, respectively, in (\ref{allup}) and (\ref{alldown}), much unlike the lepton states which are $\sim\vec S \pm \vec T$ according to (\ref{fo77}).

Just as masses, CKM matrix elements are running, i.e. dependent on the scale paramter $t=\ln \frac{E}{\mu}$ where E is the relevant energy scale and $\mu$ the renormalization scale. The running of the absolute values of the CKM matrix elements has been discussed for the 2HDM in \cite{juarez}. It turns out to be remarkably simple, at least in leading order, because it can be given in terms of one universal function h(t). 
\begin{eqnarray}
|V_{CKM}(t)| \approx  
\begin{bmatrix}
|V_{ud}(0)| & |V_{us}(0)| & \frac{|V_{ub}(0)|}{h(t)} \\
|V_{cd}(0)| & |V_{cs}(0)| & \frac{|V_{cb}(0)|}{h(t)} \\
\frac{|V_{td}(0)|}{h(t)} & \frac{|V_{ts}(0)|}{h(t)} & |V_{tb}|(0)
\end{bmatrix} 
\label{mmccrr1}
\end{eqnarray}
For the Jarlskog invariant one has
\begin{eqnarray}
J_{CKM}(t) \approx \frac{J_{CKM}(0)}{h^2(t)} 
\label{mmccrjr}
\end{eqnarray}
In the 2HDM case $h(t)$ is a moderately varying function. According to \cite{juarez} it increases by about 20\% when going from GeV to Planck scale energies.

Using (\ref{wck41111}) and (\ref{sm15}) I have calculated the CKM elements at high energies and then extrapolated them back to GeV energies according to (\ref{mmccrr1}). I obtain the matrix $|V_{CKM}|$ of absolute values
\begin{eqnarray} 
\begin{bmatrix}
   0.974 & 0.224 & 0.0035  \\
    0.224 & 0.973 & 0.044  \\
      0.0080 &  0.043 & 0.9991  \\
\end{bmatrix}
vs.
\begin{bmatrix}
   0.9734-0.9740 & 0.2235-0.2251 & 0.00362-0.00402  \\
    0.217-0.225 & 0.969-0.981 & 0.0394-0.0422  \\
      0.0083-0.0088 &  0.0404-0.0424 & 0.985-1.043  \\
\end{bmatrix}
\label{ckm1aa}
\end{eqnarray} 
The numbers look reasonable, as compared to the phenomenological values\cite{pdg}, and show the correct hierarchy and orders of magnitude. However, the theoretical uncertainty from the scale evolution is large and difficult to estimate, in particular concerning quark mass values near the Planck scale. For example, $\epsilon_{1}$ accommodates the Cabbibo angle correctly, whereas the `23'-matrix elements $|V_{ts}|$ and $|V_{cb}|$ tendencially come out too large, while the `13'-elements $|V_{ub}|$ and $|V_{td}|$ are typically too small. These deviations may seem being just 2$\sigma$ effects, but as stressed before the theoretical error from the quark mass evolution is extremely difficult to handle.

Similarly, concerning the Jarlskog invariant one obtains $J_{CKM}=0.000027$, a bit small when compared to the observed value\cite{pdg} $J_{CKM}=(3.00+0.15-0.09) \times 10^{-5}$.

\section{Supplementary material III: The role of Higgs and gauge bosons in the calculation} 


Concerning the Higgs and gauge boson contributions, one should analyze combined excitations of one of the isospin vectors $\vec S_1:= \vec Q_L$ on a tetrahedron 1 and another isospin vector $\vec S_2:= \vec Q_R$ on a neighboring tetrahedron 2. 
It is further assumed that all the other isospin vectors in the 2 tetrahedrons vibrate with $\vec S_1$ and $\vec S_2$ in a synchronous way. 
Left and right density vibrations $n_1$ and $n_2$ are defined in an analogous fashion in terms of $n_L$ and $n_R$ given in (\ref{pm11gg92}).

The tetrons which build up the vectors $\vec S_1$ and $\vec S_2$ are denoted by 
\begin{eqnarray}
\Psi_1=
\begin{bmatrix}
\delta D_1 \\
\langle U \rangle + \delta U_1 \\
\end{bmatrix}
\qquad \qquad
\Psi_2=
\begin{bmatrix}
\delta D_2 \\
\langle U \rangle + \delta U_2 \\
\end{bmatrix}
\label{pm112}
\end{eqnarray}
where $\Psi_1$,  $\delta D_1$ and $\delta U_1$ are lefthanded fields and $\Psi_2$,  $\delta D_2$ and $\delta U_2$ righthanded ones.
Such an ansatz is always allowed since one is just writing the fields as a non-chiral vev $\langle\Psi_{1,2} \rangle=(0,\langle U \rangle)$ corresponding to a state where the isospin vectors $\vec S_{1,2}$ are aligned in the ground state and point in the z-direction, plus a rest, where the `rest' consists of vibrations $\delta$ around this ground state.

Considering vibrations of tetrons 1 and 2 in (\ref{pm112}), there are altogether 8 vibrational degrees of freedom. Quite in general 4 of the 8 vibrational eigenstates are given by
\begin{eqnarray}
\delta \, \mathfrak{Re} (D_1- D_2), \quad
\delta \, \mathfrak{Im} (D_1- D_2), \quad
\delta \, \mathfrak{Re} (U_1- U_2), \quad
\delta\, \mathfrak{Im} (U_1- U_2)
\label{pm114}
\end{eqnarray}
whereas the other 4 combinations (with the plus sign) do not play any physical role in an environment of many tetrahedrons.

So the next step is to consider Heisenberg interactions of 2 vectors $\vec S_1$ and $\vec S_2$ sitting in neighboring tetrahedrons and interacting via an iso-ferromagnetic Hamiltonian. The boson masses will then arise from {\it inter}-tetrahedral isospin interactions (while quark and lepton masses are due to {\it inner}-tetrahedral ones).

Let me start with the spin-1 fields and discuss the spin-0 case later. The expressions (\ref{pm114}) are associated to vibrations of $\vec S_{x}$ , $\vec S_{y}$, $n$ and $\vec S_{z}$, respectively, to be interpreted as the SM fields $W_x$, $W_y$, $B$ and $W_z$. The physical states $W^\pm$ then correspond to $\delta (D_1-D_2)$ and $\delta (D_1-D_2)^\dagger$, and photon and Z-boson to a mixture of the U-vibrations, as explained below after (\ref{cvev1}).

In contrast to the quark and lepton mass calculation\cite{bodoprep} one should start here from the Hamiltonian and not from the equations of motion, because density contributions can then be included more easily. The relevant expression due to isomagnetic exchange is purely of `ferromagnetic' type, because 2 isospin vectors of neighboring tetrahedrons tend to align, and as discussed before there is no contribution from DM-interactions.
\begin{eqnarray}
H_{inter}^{(1)}=-\frac{1}{\Lambda^2}  [ g^2 \vec S_1 \vec S_2 + g'^2  n_1 n_2 ] \sim c_W^2 \vec S_1 \vec S_2 + s_W^2 n_1 n_2
\label{pm1137}
\end{eqnarray}
where $inter$ refers to the inter-tetrahedral interactions and the superscript $(1)$ to the spin-1 case, i.e. to the gauge bosons. $g$ and $g'$ are the SM gauge couplings and $s_W$ and $c_W$ sine and cosine of the Weinberg angle. 

In order to derive (\ref{pm1137}) one should remember 
that the isospin vectors $\vec S$ generate the Lie group of isospin rotations which in the SM corresponds to the $SU(2)_L$ gauge symmetry with coupling $g$ while the tetron densities generate the SM U(1) gauge symmetry with coupling $g'$.

$\Lambda$ is a new energy scale whose significance will be discussed later after (\ref{pm1vf}). It turns out that as far as the SM is concerned, $\Lambda$ can be absorbed in a rescaling of the tetron fields. This means that the values of $g$ and $g'$ effectively determine (and are determined by) the strength of the interaction between isospinors in neighboring tetrahedrons.


$H_{inter}^{(1)}$ is reminiscent of the negative $-D(x_1-x_2)^2\equiv -Dx^2$ of the potential of a coupled harmonic oscillator, corresponding to a parabola in the eigencoordinate $x$. For a SSB to occur, however, an additional positive contribution $\sim x^4$ is needed in the potential
\begin{eqnarray}
V(x)=-Dx^2+h x^4
\label{x1m114}
\end{eqnarray}
to obtain a minimum. 

Note that such a quartic term is not included in (\ref{pm1137}). Its existence has to be assumed, but for determining the masses of the excitations knowledge of its precise form is actually not needed. The point is that the masses can be given in terms of the quadratic coefficients alone, because they are determined by the curvature at the minimum of the potential. This curvature turns out to be $+2D$ in the case of (\ref{x1m114}) and so does not to depend on h but only on D. The situation is the same in the case of (\ref{pm1137}) and in fact also in the Higgs potential case where $m_H^2=2\mu^2$ does not depend on the $\Phi^4$ coupling value.

One can now work out (\ref{pm1137}) with the help of (\ref{pm112}) and identify the masses from the terms quadratic in $\delta$. More precisely, the coefficient of $\sim [\delta \, \mathfrak{Re} (D_1- D_2)]^2 +[\delta \, \mathfrak{Im} (D_1- D_2)]^2$ yield the mass squared of W$^\pm$. One obtains the SM result for the W-mass $m_W^2=g^2v_F^2/4$ under the condition that the order parameter, i.e. the Fermi scale $v_F$ is given by
\begin{eqnarray}
\frac{v_F^2}{2}=\frac{|\langle U \rangle|^2}{\Lambda}
\label{pm1vf}
\end{eqnarray}
In order to obtain the mass for the Z-boson and also the correct mixing of the $W_z$ and $B$ boson field one has to allow for a complex vev 
\begin{eqnarray}
\langle U \rangle =  |\langle U \rangle| \exp {(i \theta_W)} 
\label{cvev1}
\end{eqnarray}

Evaluation of (\ref{pm1137}) shows that the phase of $\langle U \rangle$ 
must indeed be chosen to be the Weinberg angle $\theta_W=\arctan (g'/g)$, 
because this leads to one massive combination $Z=W_z c_w-B s_W$ and 
one massless combination $A=W_z s_w+B c_W$, with the SM 
result for the Z-mass  $m_Z^2=(g^2+g'^2)v_F^2/4$ being recovered.

At first sight $\Lambda$ according to (\ref{pm1137}) seems to crucially affect the strength of isomagnetic exchange. However, according to (\ref{pm1vf}) the 'strength' of the electroweak SSB is determined by a ratio involving $\langle U \rangle$ and $\Lambda$, and one can actually grossly absorb all effects of $\Lambda$ in a redefinition of the tetron fields $\Psi\rightarrow \Psi /\sqrt{\Lambda}$. This rescaling can be interpreted as reducing the `high' Planck scale values of the tetron fields to the `low' level of the Fermi scale. 
Thus, from the very perspective of the SM, the gauge couplings $g$ and $g'$ alone determine (and are determined by) the strength of the isomagnetic exchange.

So, from the viewpoint of the SM, the absolute values of $|\langle U \rangle|^2$ and $\Lambda$ are not relevant, but only their ratio $v_F$. On the other hand, from the viewpoint of the tetron model the values of $|\langle U \rangle|$ and $\Lambda$ each have a separate physical meaning, and so have the ratios $g^2/\Lambda^2$ and $g'^2/\Lambda^2$, because these quantites according to (\ref{pm1137}) correspond to the iso-ferromagnetic couplings of tetron isospins and should be calculable from first principles, i.e. from the form of the fundamental tetron interactions. 

If one thinks more closely, only the ratio $g/g'$ (i.e. the Weinberg angle) and the Fermi scale $v_F$ are related to the isomagnetic exchange forces, while the third independent parameter, which is given by the fine structure constant, relates more to the direct (as opposed to exchange) interactions of tetrons, and in fact to gravity\cite{bodohiggs,bodotalk}.

Within the tetron approach it is natural to assume that the Weinberg angle with measured value $(28.70\pm 0.05)^{\circ}$ is related to the geometry of the tetrahedron - in some way or other. 
In the following I want to suggest 2 possibilities:\\
(i) `Hybridization' of isospin-1 vibrations: The 3 orthogonal directions in which $\vec S_{x}$ , $\vec S_{y}$ and $\vec S_{z}$ vibrate do not fit well into the tetrahedral structure of 4 tetrons and therefore the states `hybridize' with the radially symmetric vibration of the density.
For the simplified model considered here, with $\langle \vec S \rangle$ along the z-direction, this amounts to a mixture of the $n$ and $S_z$ vibrations with a mixing angle of exactly $\theta_W=30^{\circ}$ and a corresponding relative magnitude of $g/g'=\sqrt{3}$.\\
(ii) Enforcement of the Broglie-Bohr quantization condition: The angle between any 2 isospin vectors in a tetrahedron is given by $\theta_T=\arccos(-\frac{1}{3}) \approx 109.5^{\circ}$ and in geometry is usually called the `tetrahedral angle'. Thus in order that the complete wave function corresponds to a standing wave around the 4 corners of a tetrahedron, each tetrahedral corner must contribute $\theta_T$. This means the left and right components on each site must contribute $\theta_T/2$ each, and since tetrons are fermions this amounts to a phase $\theta_W=\theta_T/4\approx 27.4^{\circ}$ in the tetron wave function.

We now turn to the spin-0 states of the SM. They constitute the complex Higgs doublet of the form
\begin{eqnarray}
\Phi=\frac{1}{\sqrt{2}} \exp  ( \frac{i}{v_F}\vec\tau \vec\xi) 
\begin{bmatrix}
0 \\
v_F+H \\
\end{bmatrix}
\label{gue1}
\end{eqnarray}\\
where $\vec\xi$ is the triplet of Goldstone bosons and H the physical Higgs field. As explicit from (\ref{gue1}), the $\xi$ fields can be gauged to zero by an appropriate SU(2) transformation. This means, although the concept of Goldstone bosons is crucial to understanding symmetry breaking in the Standard Model, there are no physical Goldstone bosons in the observed spectrum.

How does this translate to the microscopic theory? The isospin vibrations (\ref{pm114}) can in principle generate spin-0 fields $\xi_x$, $\xi_y$, $H$ and $\xi_z$. Since spin-0 and spin-1 wave functions are different in the base space, the modes for $\xi_x$, $\xi_y$, $H$ and $\xi_z$ are different from the gauge boson modes $W_x$, $W_y$, $B$ and $W_z$, and due to this difference the exhange integrals and accordingly the couplings appearing in the iso-magnetic Hamiltonian will be different as well. Instead of (\ref{pm1137}) one has
\begin{eqnarray}
H_{inter}^{(0)}=-\frac{\mu^2}{\Lambda^4}    n_1 n_2
\label{pm1713}
\end{eqnarray}
where $\Lambda$ is as above and $\mu^2$ the parameter well-known from the Higgs potential leading to a Higgs mass of $m_H^2=2\mu^2$. The missing Heisenberg contribution $\sim \vec S_1 \vec S_2$ in (\ref{pm1713}) makes explicit that there are actually no vibrations which would correspond to particles $\xi_x$, $\xi_y$ and $\xi_z$.


\begin{thebibliography}{99}
\bibitem{bodomasses} B. Lampe, Int. J. Mod. Phys. A31 (2016) 1650115, 1650116.
\bibitem{moriya} T. Moriya, Phys. Rev. 120 (1960) 91.
\bibitem{maki} Z. Maki, M. Nakagawa and S. Sakata, Progr.Theor. Phys. 28 (1962) 870.
\bibitem{giganti} C. Giganti, S. Lavignac and M. Zito, Progr. Part. Nucl. Phys 98 (2018) 1.
\bibitem{jarls} C. Jarlskog, Z. Phys. C29 (1985) 491.
\bibitem{pdg} Particle Data Group, Prog. Theor. Exp. Phys. 8 (2022 and 2023 update) 083C01.
\bibitem{jarl2}  A. R. Ellis, K. J. Kelly and S. W. Li, Phys. Rev. D 102 (2020) 115027.
\bibitem{bodohiggs} B. Lampe, Progr. of Phys. 69 (2021) 2000072.
\bibitem{bodotalk} B. Lampe in Proc. of the MG16 Meeting on General Relativity, doi.org/10.1142/13149 (2023) 999.
\bibitem{cab} N. Cabibbo, Phys. Rev. Lett. 10 (1963) 531.
\bibitem{bodoprep} B. Lampe, Fortsch. Phys. 72 (2024) 5, 2300258.
\bibitem{runnmass} K. Bora, arXiv:1206.5909 [hep-ph] (2012).
\bibitem{juarez} W. Juarez et al., Phys. Rev. D66 (2002) 116007. 
\end{thebibliography}
\end{document}